\documentclass[iop]{emulateapj} \usepackage{latexsym} \usepackage{makeidx}
 \usepackage{amsmath} \usepackage{color} \usepackage{upgreek}
\usepackage{morefloats} \usepackage{lineno} \usepackage{setspace}
\usepackage{graphicx}

\shorttitle{Clouds on the hot Jupiter HD189733b: constraints from the reflection
spectrum} \shortauthors{Barstow et al.} \begin{document}

\title{Clouds on the hot Jupiter HD189733b: constraints from the reflection spectrum}
\author{J.~K. Barstow, S. Aigrain, P.~G.~J. Irwin, T. Hackler and L.~N. Fletcher}
\affil{Department of Physics, University of Oxford, Oxford, UK}, \author{J.~M.
Lee} \affil{Institute for Theoretical Physics, University of Z\"urich, CH-8057
Z\"urich, Switzerland} \and \author{N.~P. Gibson} \affil{European
Southern Observatory, 85748 Garching bei M\"unchen, Germany}\email{jo.barstow@astro.ox.ac.uk}

\label{firstpage}

\begin{abstract} The hot Jupiter HD 189733b is probably the best studied of the known
extrasolar planets, with published transit and eclipse spectra covering the near UV
to mid-IR range. Recent work on the transmission spectrum has shown clear evidence
for the presence of clouds in its atmosphere, which significantly increases the model
atmosphere parameter space that must be explored in order to fully characterise this
planet. In this work, we apply the NEMESIS atmospheric retrieval code to the recently
published \textit{HST}/STIS reflection spectrum, and also to the dayside thermal emission
spectrum in the light of new \textit{Spitzer}/IRAC measurements, as well as our own
re-analysis of the \textit{HST}/NICMOS data. We first use the STIS data to place some constraints on the nature of the cloud on HD
189733b, and explore solution degeneracy between different cloud properties and the
abundance of Na in the atmosphere; as already noted in previous work, absorption due to Na plays a significant role in
determining the shape of the reflection spectrum. We then perform a new retrieval of
the temperature profile and abundances of H$_2$O, CO$_2$, CO and CH$_4$ from the
dayside thermal emission spectrum. Finally, we investigate the effect of including
cloud in the model on this retrieval process. We find that the current quality of
data does not warrant the extra complexity introduced by including
cloud in the model;
however, future data are likely to be of sufficient resolution and signal-to-noise
that a more complete model, including scattering particles, will be
required.\end{abstract}


\maketitle

\section{Introduction} Since its discovery in 2005 \citep{bouchy05}, the hot Jupiter
HD 189733b has been repeatedly observed as it transits and is eclipsed by its parent
star, leading to excellent coverage in both its transmission and
eclipse spectra from the visible to mid-infrared.
This has resulted in HD 189733b being probably the best characterised of all the
known exoplanets; it is known from the observed slope of the
reflectance spectrum that this unresolved planet would appear a deep shade
of blue \citep{evans13}, a fact that testifies to the power of the transit
spectroscopy technique. The transmission spectrum presented by \citet{pont13} shows
clear evidence for haze or cloud in the atmosphere of this planet, making it the
first transiting planet outside the solar system that is known to be cloudy. \citet{lee12} and
\citet{line12} analyse the dayside spectrum from secondary eclipse observations, and
retrieve the temperature-pressure profile and abundances of H$_2$O, CO$_2$, CO and
CH$_4$. \citet{knutson12} use \textit{Spitzer}/IRAC phase curves to investigate the
longitudinal temperature variability, and \citet{dewit12} combine the phase curves
with an analysis of the ingress/egress shape in eclipse to place further constraint
on spatial variability.

The recent albedo spectrum from \textit{HST}/STIS obtained by \citet{evans13}
provides an opportunity to investigate the cloud structure on the dayside. Unlike the
transmission spectrum investigated by \citet{pont13} and \citet{lee13} which probes
the limb of the exoplanet amosphere, the dayside reflection spectrum has near-nadir
geometry, and so it is sensitive to deeper regions of the atmosphere. If the cloud is
similar at the terminators and on the dayside, we can use this to further constrain
its properties. Alternatively, we may see an entirely different cloud layer on the
hotter dayside from that observed at the terminator. The albedo spectrum is useful
for placing constraint on the cloud, as scattering particles have a significant
effect on the optical reflectivity of an atmosphere. We expect there
to be fewer gaseous
absorbing species in the visible part of the spectrum than in the infrared; distinct
absorption features of Na and K are seen in the transmission spectrum, but otherwise
cloud appears to be the dominant opacity source in this region
\citep{pont13}. We also expect little, if any, thermal contribution
from the planet itself at these wavelengths due to its temperature. 

In this work, we use the  Non-linear optimal Estimator for MultivariateE spectral
analySIS (NEMESIS) software \citep{irwin08} to calculate synthetic spectra using a
simple cloudy model atmosphere for HD 189733b, including multiple scattering. We vary
cloud parameters and Na volume mixing ratio (VMR) and compare the resultant spectra
to the STIS measurement; calculating the $\chi^2$ goodness-of-fit parameter allows us
to determine the region of model parameter space that provides the best match. We
then use a subset of our best-fitting models to examine the effect of clouds on our
ability to accurately retrieve temperature and molecular abundances from the infrared
dayside emission spectrum, following the work of \citet{lee12}.

\subsection{NEMESIS} We use the NEMESIS spectral retrieval tool to produce forward
models (predicted spectra for a range of model atmospheres) for comparison with the \textit{HST}/STIS spectra, and also to retrieve the
atmospheric state from the thermal emission spectrum as in \citet{lee12}. NEMESIS was
developed by \citet{irwin08} for atmospheric retrieval of solar system planets, and
has since been extended to enable the same analysis for observations of
transiting and directly imaged extrasolar planets \citep{lee12,lee13a,barstow13b}. NEMESIS is an Optimal Estimation
retrieval model \citep{rodg00} and uses a correlated-k radiative transfer model
\citep{lacis91,goodyyung}. NEMESIS is not a radiative
  equilibrium model; instead, it simply uses the atmospheric model
  provided to compute the incoming and outgoing radiative flux. In an irradiated
  case, it will compute the incoming and scattered/reflected flux from the star,
  but it will not take into account the heating effect of the incoming stellar
  flux on the atmosphere. For the multiple scattering runs, NEMESIS
  uses beams over a user-specified number of zenith angles; azimuthal
  dependence is accounted for using Fourier decomposition. For more
  details of the scattering calculation in this work, see Section~\ref{multscat}.

We use the same line and collision-induced absorption data as \citet{lee12},
\citet{lee13} and \citet{barstow13}. A list of sources for
absorption line data is given in Table~\ref{linedata}. H$_2$-He
collision-induced absorption data are taken from the models of
\citet{borysow89,borysowfm89,borysow90,borysow97} and \citet{borysow02}. The reference
stellar spectrum is taken from the model set made available by
Kurucz\footnote{http://kurucz.harvard.edu/stars/HD189733/}, and the
stellar radius is taken from \citet{ellyn08}. We use the same
planetary mass and radius as \citet{lee13}, and a H$_2$:He ratio of
9:1. As found by \citet{lee12}, the precise value of this does not
have a large effect on secondary eclipse retrievals. 

\begin{table} \centering \begin{tabular}[c]{|c|c|} \hline Gas & Source\\ \hline
H$_2$O & HITEMP2010 \citep{roth10}\\ CO$_2$ &  CDSD-1000 \citep{tash03}\\ CO &
HITRAN1995 \citep{roth95}\\ CH$_4$ & STDS \citep{weng98}\\ Na & VALD
\citep{heiter08}\\ K & VALD \citep{heiter08}\\ \hline \end{tabular} \caption{Sources
of gas absorption line data.\label{linedata}} \end{table}

\subsubsection{Multiple scattering calculations}\label{multscat} In
order to reproduce an accurate reflection
spectrum in the presence of optically thick cloud, it is necessary to
include multiple scattering as many scattering events are likely. NEMESIS uses the matrix
operator algorithm of \citet{plass73} for multiple scattering calculations, where the
zenith angle integration is achieved using a 5-point Gaussian-Lobatto quadrature
scheme and azimuth angle integration is achieved through Fourier decomposition, with
the necessary number of Fourier components determined by the stellar and emission
zenith angles. The analytical disc-averaged integration scheme used in \citet{lee12}
for eclipse spectra is not used here as it is not applicable to scattering
situations; instead, we represent the disc average for the synthetic STIS spectra by
running multiple scattering calculations with the stellar zenith angles set
to each of the 5 Gaussian-Lobatto quadrature angles and the azimuth angle set for
back-scattering, since during secondary transit the observer is
located in the same direction as the star and the stellar zenith angle
is equal to the emission angle. The disc-average is then determined using a weighted average of
these 5 different calculations, assuming that the atmospheric conditions are the same
at all points on the disc. Retrievals including multiple scattering are
computationally expensive. Therefore, we anticipate that the majority of the thermal
emission retrieval calculations in the future will still be performed
with an extinction-only approximation using the disc integration described by
\citet{lee12}. However, we also tested the effect of including multiple scattering,
to test the sensitivity of the retrieval to differences in our modelling approach.
The 5-angle approach used to calculate the STIS synthetic spectra is still too
time-consuming for the emission spectrum retrieval, so in this case we approximate
further by calculating the spectrum for a stellar zenith angle of 45$^{\circ}$ only,
which represents the average angle of weighting function used to compute the
disc-averaged spectrum since
$\bar{R}=\int_0^{\frac{\pi}{2}}\!2R(\theta)\mathrm{sin}(\theta)\mathrm{cos}(\theta)\,
\mathrm{d}\theta=\int_0^{\frac{\pi}{2}}\!R(\theta) \mathrm{sin}(2\theta)
\,\mathrm{d}\theta$.

The scattering phase function of the particles is calculated using Mie theory and
approximated by the Henyey-Greenstein parameterisation \citep{hg41}, as in previous
work on planetary clouds (e.g. \citealt{irwin09}, \citealt{barstow12}); expected
deviations from the true phase function are small compared with the errors on the
observed spectrum, so we consider this approximation to be valid. We use the double-peaked version of the phase function:

$P(\theta)= \frac{1}{4\pi}\left[f\frac{1-{g^2_1}}{(1+{g^2_1}-2{g_1}\mu)^{3/2}}+(1-f)\frac{1-{g^2_2}}{(1+{g^2_2}-2{g_2}\mu)^{3/2}}\right]$,

which represents the phase function as the sum of forward and backward scattering peaks. Here, $\mu$ is the cosine of the scattering angle, $g_1$ and $g_2$ are scattering asymmetry parameters for the forward and backward peaks respectively, and $f$ is the fractional contribution of the forward peak to the total phase function. For smaller particles approaching the Rayleigh scattering limit, the parameter $f$ is close to 0.5 and there are approximately equal contributions from the forward and backward scattering peaks; as the particle size increases relative to the wavelength of light the value of $f$ increases and the scattering becomes more asymmetric. However, in no case is the scattering completely isotropic.

We use the enstatite refractive index values of \citet{scott96} and the
MnS values from \citet{huff67} in our calculations of the scattering parameters. It is worth noting that, as we model realistic particles, they absorb as well as scatter incident radiation. The fraction of light that is absorbed rather than scattered is dependent on the particle size and also on the composition of the particles. In general, more of these realistic particles would be required than idealised, perfectly scattering particles to produce an equivalent planetary albedo in an atmospheric model, since for realistic particles some of the incident radiation is absorbed by the particles rather than scattered back to space. The single scattering albedo is also wavelength dependent, and this can therefore affect the shape of modelled planetary albedo and thermal emission spectra.

\section{Data} The \textit{HST}/STIS data used in this work are presented in
\citet{evans13}. We use the tabulated eclipse depths binned in 6 channels, spanning
the wavelength range 290---570 nm, to place constraint on the cloud properties for HD
189733b.

We also investigate the impact of including clouds on retrievals of temperature
structure and atmospheric composition from the available thermal emission spectra. We
use the same data as \citet{lee12}, with the exception that we include the new values
for the \textit{Spitzer}/IRAC 3.6 and 4.5 $\upmu$m points presented in
\citet{knutson12}. Data used as in \citet{lee12} are the \textit{Spitzer}/IRS spectrum from
\citet{grillmair08}, the \textit{Spitzer}/IRS broadband point
originally measured by \citet{deming06} and the broadband \textit{Spitzer} MIPS and
\textit{Spitzer}/IRAC 5.8/8.0 $\upmu$m points from
\citet{charb08}. 

We also include a reanalysis of the \textit{HST}/NICMOS observations of
\citet{swain09}. \citet{gibson11} re-analysed a number of exoplanet transmission
spectra obtained with NICMOS, including that of XO-1b, originally published by
\citet{tinetti10}. This re-analysis showed that the shape of the NICMOS spectra is
highly sensitive to the assumptions made about the instrumental systematics, and that
the errors may have been under-estimated in the original publications. This was recently
confirmed by new observations of XO-1b in transit with \textit{HST}/WFC3
\citep{deming13}, which are less affected by systematics, and showed significant
discrepancies with the original NICMOS dataset. We expect that the NICMOS emission
spectrum of HD189733b may suffer from the same problems as the aforementioned
transmission spectra, but it is the only dataset published to date in the crucial
near-infrared wavelength range. We therefore re-analysed it using a method proposed
by \citet{gibson12}, which makes minimal assumptions about the nature of the
systematics. This re-analysis, which we describe briefly in Section 2.1, leads to
larger, but somewhat more robust error bars, and thus more conservative results for
the retrieval analysis.

We omit the ground-based K- and L-band data of \citet{waldmann12}, as these data
contain an extremely strong emission feature at 3.3 $\upmu$m which is attributed to
non-Local Thermodynamic Equilibrium (non-LTE) processes; modelling this feature is
therefore beyond the limits of our code. The attribution of this
feature as resulting from the atmosphere of HD 189733b has also been disputed by
\citet{mandell11} as these authors did not detect it; they suggest it
may be a result of telluric absorption. 
This conflict is so far unresolved and as a result we do not include
these data in our analysis.

\subsection{Reanalysis of NICMOS emission spectrum} \label{nicmosanalysis} We
re-analysed the NICMOS emission spectrum presented in \citet{swain09} using the method
presented by \citet{gibson12}. We briefly outline the process here, but refer the
interested reader to \citet{gibson12} for further information on the data
reduction and the details of the systematics modelling method. We model the secondary
eclipse and systematics simultaneously in each of 18 wavelength channels, using
Gaussian Processes (GP) to model the dependence of the systematics on the
instrumental parameters which are known to affect them: the orbital phase of HST
$\phi$ and its square $\phi^2$; the $x$ and $y$-position of  the spectral trace on
the detector, its width $w$ and its angle $\theta$ relative to the
x-axis. The ephemeris, planet-to-star radius ratio, inclination and
system scale ($a/R_{\star}$) were fixed to the values obtained by
\citet{winn06}, These are also consistent with the more recent
determination of these parameters by \citet{southworth08}. The eccentricity was assumed to be zero, as there is no evidence in the literature for a non-zero eccentricity. We used three Markov-Chain Monte Carlo (MCMC) chains of 40,000 steps each to
marginalise over all the parameters except the eclipse depth in each channel.

Figure 1 shows the spectrum obtained after re-analysis, with the original from \citet{swain09} for comparison. The two versions of the spectrum are largely similar in
shape, but our error bars are about a factor of 3 larger, which is consistent with
the findings of \citet{gibson12}. The wavelength range probed by NICMOS is
particularly sensitive to the gaseous abundances on the dayside of HD 189733b \citep{lee12}, so the increased error bars will impact the constraints we can extract
from the thermal emission spectrum. The spectrum and $1\sigma$ error bars are
provided in numeric form in Table 2.

We note that the choice of which instrumental parameters to include in the
systematics model for NICMOS affects both the shape of the resultant spectrum and the
errors, whether using traditional linear de-correlation methods \citep{gibson11}
or our GP approach. However, in a GP model, the length scale $\eta$ associated with
each parameter gives an indication of the relevance of that parameter (see \citealt{gibson12} for details). In the present case, the best-fit length scales for all the
instrument parameters we tried including in the model were found to be of the same
order of magnitude, which is why we included them all in the final model.
Nonetheless, this highlights the fact that the NICMOS results should be viewed with
caution, as the spectrum is extremely sensitive to the details of the systematics
removal.

\begin{figure} \includegraphics[width=0.5\textwidth]{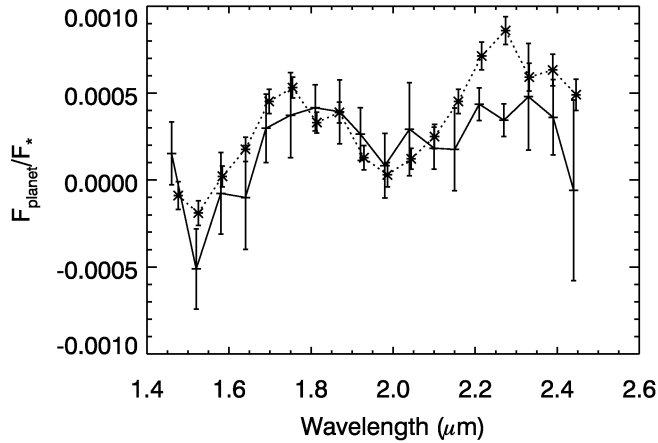} \caption{The NICMOS
spectrum as analysed by Swain et al. (2009; stars, dotted line) and in this work
(crosses, solid line). We find that the spectrum has a similar shape but the feature
amplitude is decreased and the error bars are increased.\label{nicmos}} \end{figure}

The reanalysed spectrum and 1$\sigma$ errors are provided in Table~\ref{nicmosdat}.

\begin{table} \centering \begin{tabular}[c]{|c|c|c|} \hline Wavelength & Flux ratio
(10$^{-4}$)& Error \\ \hline 1.4638&      1.53	&1.81\\ 1.5214&    -5.11	&2.31\\
1.5790&    -0.762	&2.34\\ 1.6366&    -1.01	&2.99\\ 1.6941&      2.98	&1.96\\ 1.7518&
3.73	&2.44\\ 1.8094&      4.16	&1.33\\ 1.8670&	3.93	&1.84\\
1.9246&	2.65	&1.52\\ 1.9822&	0.834	&1.86\\ 2.0398&	2.93	&2.68\\
2.0974&	1.83	&1.20\\ 2.1550&	1.76	&2.38\\ 2.2126&	4.37	&0.941\\
2.2702&	3.44	&0.937\\ 2.3278&	4.79	&3.06\\ 2.3854&	3.61	&2.17\\
2.4430&	-0.588 &	5.20\\ \hline \end{tabular} \caption{The reanalysed NICMOS
spectrum.\label{nicmosdat}} \end{table}

\section{Cloud models} \label{cloudmodels} We use the \textit{HST}/STIS data to
investigate the range of cloud models that provide a reasonable match with HD
189733b's reflection spectrum. We use the best-fit atmospheric temperature profile
retrieved by \citet{lee12b}; the other parameters with a significant effect in the
STIS wavelength range are all related to cloud, except for the abundance of sodium.
Sodium has been detected in transmission spectra of HD 189733b
\citep{redfield08,huitson12}, but it is not possible to obtain an absolute constraint
on the sodium abundance from these
measurements. Atmospheric models by \citet{sudarsky00}
and \citet{heng13} suggest that Na should have a measurable effect on
the reflection
spectrum. The abundance is left as a free parameter in the model.

Clouds are extremely complex phenomena in radiative transfer. The resultant spectrum
for a cloudy atmosphere depends on the scattering properties of the cloud, which in
turn are affected by the composition and size of the cloud particles. The altitude
and optical depth of the cloud affect how far light from the star will penetrate the
atmosphere, which affects the fraction of the light that is absorbed before it can be
reflected back. In order to investigate the potential cloud structure in the simplest
possible way, the cloud has been modelled for a fixed set of particle sizes, optical
depths and cloud altitudes; the free parameters in our model for the
STIS spectrum are listed in Table~\ref{parvalues}, with the range of
values taken by each. The scattering efficiency and phase function for
the cloud are calculated using Mie theory with a double-peaked
Henyey-Greenstein approximation to the phase function; the scattering
properties are therefore a consequence of the size and composition of
the cloud particles.

All other parameters, including the atmospheric
temperature and H$_2$O, CO$_2$, CO and CH$_4$ VMRs, are fixed during
this analysis at the best-fit values of \citet{lee12b}; none of these gases are expected to have strong absorption
features between 0.3 and 0.6, so their only effect on the spectrum is
a small effect
on the mean molecular weight. We also include Rayleigh scattering from
H$_2$ and He, and the abundances of these gases are also fixed.

\begin{table} \centering \begin{tabular}[c]{|c|c|} \hline Variable & Values \\ \hline
Base pressure (mbar)& 1000, 100, 10, 1, uniform \\ Particle
size ($\upmu$m) & 0.01, 0.03, 0.1, 0.3, 1, 3, 10 \\ Optical depth at 0.25 $\upmu$m &
0.1, 0.2, 0.4, 0.6, 0.8, 1, 10\\ Na VMR (ppmv) & 0.5, 5, 50, 500 \\ \hline
\end{tabular} \caption{The values used for each parameter in the cloud model, and for
Na volume mixing ratio. A synthetic spectrum is generated for every combination of
these parameters, making a total of 980 models. All cloud models span
a decade in pressure, except for the uniform case - so if the base
pressure is 1000 mbar, the top pressure is 100 mbar.\label{parvalues}} \end{table}

\begin{figure} \includegraphics[width=0.5\textwidth]{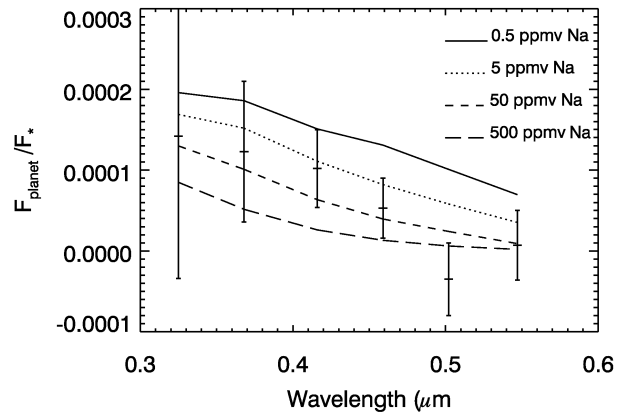} \caption{The
6-channel STIS spectrum from Evans et al. (2013), with four cloud-free models
overlaid. Models with 50 ppmv Na produce the best fit to the observed spectrum without
the requirement for cloud; models with 0.5 ppmv Na give a poor fit. 5 ppmv is
approximately solar abundance.\label{noclouds}} \end{figure}

\begin{figure*} \centering
\includegraphics[width=1.0\textwidth]{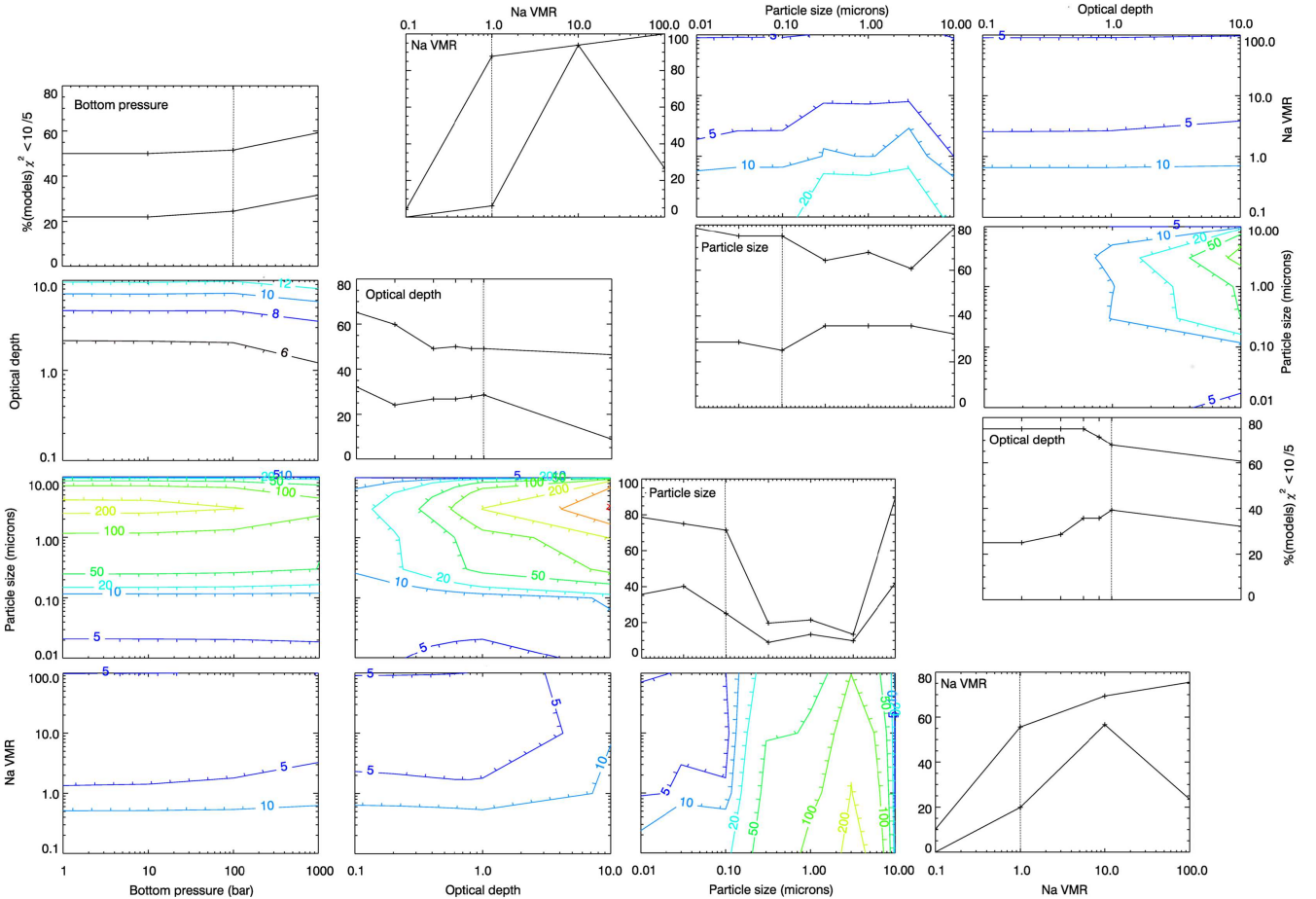} \caption{Contour plots
of $\chi^2$ goodness-of-fit parameter for (bottom left): cloud bottom pressure,
optical depth, particle size and Na volume mixing ratio; (top right): cloud optical
depth, particle size and Na volume mixing ratio for a uniformly mixed cloud. The Na
VMR is quoted as a multiplication of the solar value ($\sim$5 ppmv). Also plotted are
the percentages of models with $\chi^2$ of $<$10 (top line) and $<$5 (bottom line)
for each parameter. Each contour plot represents a 2-D
  cut through the 4-D parameter space, with two parameters varied and
  the other 
  two held fixed in each plot. The dashed lines indicate the values used for each parameter when
held fixed. \label{reflectionresults}} \end{figure*}

For most of the tests described below, we assume that the cloud particles are made of
enstatite, MgSiO$_3$ (\citealt{lecavelier08}; refractive indices used are taken from
\citealt{scott96}). This is likely to be condensed at the stratospheric temperature
on HD 189733b \citep{fortney10}, but it is not the only chemical species that may be
relevant; \citet{morley12} suggest that constituents such as MnS may also form clouds
on hot Jupiters, and we consider the impact of changing the composition of the cloud
particles in Section~\ref{composition}. Conversely, we do not expect to see
absorption due to gaseous metal oxides such as TiO in the atmosphere as Ti is likely
to be in condensate form at the expected photospheric temperature of HD 189733b
\citep{fortney10}. This may mean that any condensates present also contain Ti oxides,
as suggested by \citet{helling06}.

To test the effect of particle size, we use a range of monodisperse (single-size) cloud populations
from 10 nm to 10 $\upmu$m in radius.For the STIS wavelength range, the Rayleigh scattering limit is approached for a particle size of 10 nm, so the scattering behaviour as a function of wavelength will be similar for any particles smaller than 10 nm; the only difference would be that a larger number of smaller particles would be required to achieve the same optical depth. The nadir optical depths (referenced at 0.25
$\upmu$m) are varied between 0.1 and 10, and the cloud is located in a layer 1 decade
thick in pressure coordinates (i.e. between 1000 and 100 mbar, or 100 and 10 mbar,
etc.). We locate the cloud base at 1000, 100, 10 and 1 mbar, and also test the effect
of distributing the cloud particles uniformly throughout the atmosphere. The cloud
particle number density scale height is assumed to be the same as the pressure
scale height in all cases.

Synthetic reflection spectra are generated for all combinations of these three cloud
parameters and the Na abundance parameter, using NEMESIS. We list the values used for
each parameter in Table~\ref{parvalues}.

We also generate a series of cloud-free models to investigate the impact of changing
the sodium abundance by itself. It is clear that for a range of sodium abundances the
STIS spectrum can be reproduced without the need for clouds
(Figure~\ref{noclouds}; the effect of sodium on the spectrum is as
predicted by \citealt{sudarsky00})
but if the atmosphere in fact contains more/less sodium additional
scatterers/absorbers would be required \citep{heng13}.

For each of the cloudy models, the $\chi^2$ goodness-of-fit parameter is computed to
indicate the quality of the fit to the 6 data points in the STIS spectrum. We use
this as a means of comparing models across our parameter
space; our 4 variables are Na VMR, cloud particle radius, cloud altitude and cloud optical
depth. For a good fit, therefore, the $\chi^2$ should approach 2
($N_{data}-N_{variables}$). Contour plots of the
$\chi^2$ are shown for each parameter combination in Figure~\ref{reflectionresults},
as well as the number of models for each value of each single parameter with a
$\chi^2$ of less than 10 and less than 5. The contour plots
  show two-dimensional cuts through the four-dimensional parameter
  space, with two parameters varying in each plot and the other two
  held fixed. The values for each fixed parameter are indicated by the
  dashed lines in the histogram plots. This indicates that the best-fit region of
parameter space occurs mostly for either small particles ($<0.3$ $\upmu$m in size) or
10 $\upmu$m particles, 10$\times$ solar Na (50 ppmv) and optical depths of 1 or less.
The preference for small particles (0.01---0.1 $\upmu$m) is consistent with the
findings of \citet{lecavelier08}.

\begin{figure} \includegraphics[width=0.5\textwidth]{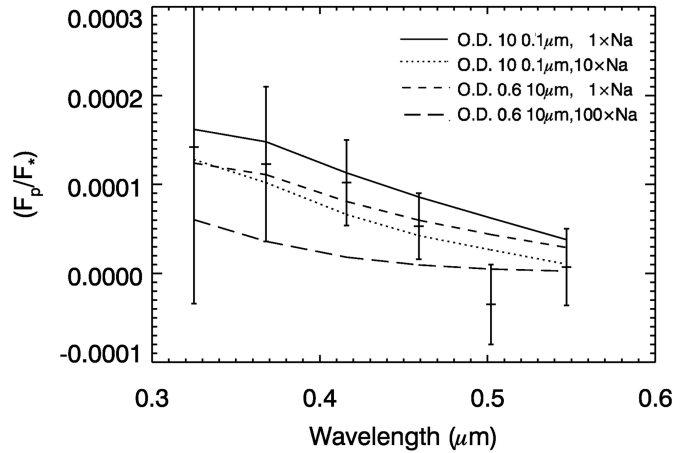}
\caption{Spectra for a selection of cloud models with different properties and Na
VMRs. The 0.1 $\upmu$m models are uniformly distributed in altitude, and the 10
$\upmu$m models are between 1 and 0.1 mbar. The 0.1 $\upmu$m model with 10$\times$Na
and the 10 $\upmu$m model with 1$\times$Na fit the spectrum well ($\chi^2<$5),
whereas the other two models do not.\label{cloudspectra}} \end{figure}

\begin{figure} \centering\includegraphics[width=0.35\textwidth]{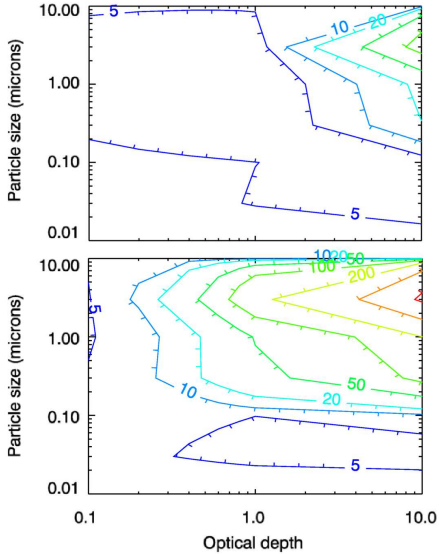} \caption{$\chi^2$
contour plots for 100$\times$ solar Na and a cloud from 1000---100 mbar (top) and
100---10 mbar (bottom). The best-fitting models for the first case occur for an
optical depth of 1 and a particle size between 0.3 and 3 $\upmu$m; for the second
case, the best-fit model occurs for an optical depth of 10 and a particle size of
0.03 $\upmu$m. \label{100xNa}} \end{figure}

A subsolar sodium abundance (0.1$\times$, or 0.5 ppmv) is insufficient to reproduce
the spectrum for the cases we test here. Examples of models with
different cloud properties and Na VMRs are shown in
Figure~\ref{cloudspectra}; models with 5---500 ppmv Na could all produce
an adequate fit to the spectrum depending on the assumed cloud
properties, but a model with 0.5 ppmv Na does not produce a
sufficiently large absorption feature at wavelengths $>$0.5 $\upmu$m. In general,
the cloud altitude is the model parameter with the smallest effect, except for the case where
the sodium abundance is 100$\times$ solar (Figure~\ref{100xNa}). The degeneracy
between sodium abundance and the cloud properties becomes complex when the sodium
abundance is high, because the cloud must make the atmosphere optically thick at
precisely the right altitude in order for the model to fit. For 50 ppmv sodium a
cloud-free model can fit the spectrum with a $\chi^2$ of less than 5
(Figure~\ref{noclouds}), but the increase in sodium abundance means that an
additional reflective component is required in the model atmosphere
for 500 ppmv.

The data do not allow us to distinguish between the
  finite deck-type models, where the cloud is confined to a certain
  pressure range, and models where the cloud particles are uniformly
  distributed throughout the atmosphere as for many cases we can obtain a good fit using
  either (Figure~\ref{reflectionresults}). Based on the results of
  \citet{parmentier13}, a uniform cloud of particles up to 0.1
  $\upmu$m in size is plausible; we may expect a cloud formed of
 larger particles to be located lower in the atmosphere
  due to sedimentation and rain out. For particle sizes of up to 0.1
  $\upmu$m, the uniform cloud model contains fewer assumptions (as we
  do not define a cloud base or cloud top pressure) and
  Occam's razor suggests that this should be preferred for any given
  values of Na VMR, particle radius and optical depth, where the quality
  of fit is equal.

\citet{heng13} also compare a set of cloud models to the \textit{HST}/STIS spectrum,
and find similar degeneracy between Na abundance and
cloud. They find that the abundance of Na relative to
  the abundance of cloud particles decreases by a factor 10$^5$ if the
  cloud particle size decreases by a factor 10 from 100 nm to 10
  nm. For sub-$\upmu$m
  particles, we find a similar result. The best fit Na VMR, and
  therefore the number of Na atoms, is roughly
  constant as a function of particle size if optical depth and cloud
  altitude are fixed (Figure~\ref{reflectionresults}), and we require a factor 10$^5$ increase in the
  number of cloud particles as the size decreases from 100 nm to
  10 nm to maintain a fixed optical depth. As in the work of \citet{heng13}, we find that the shape of the reflection spectrum is largely dictated by the particle size and the Na VMR.

\subsection{K, TiO and VO} 
\begin{figure} \includegraphics[width=0.5\textwidth]{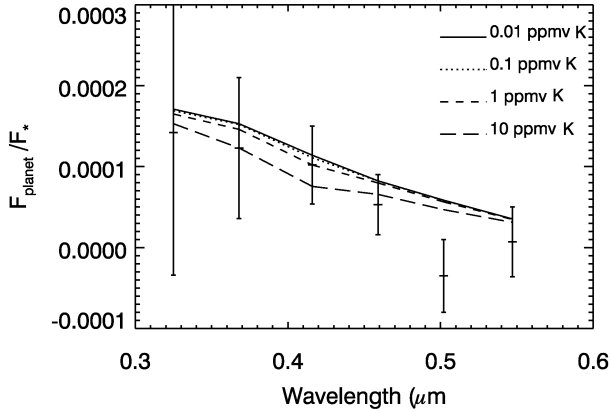} \caption{Synthetic,
  cloud-free STIS
spectra containing different abundances of K. Varying the amount of K in the model
does not significantly affect the spectrum.\label{potassium}} \end{figure}

So far, we have only examined the effect of Na on the
visible spectrum, and have ignored other species. We include K in the model
atmosphere (0.1 ppmv), and Figure~\ref{potassium} shows the effect of varying the
abundance over 4 orders of magnitude for a cloud-free model. It can be seen that K
has a much smaller effect on the shape of the spectrum than Na
(Figure~\ref{noclouds}), and the magnitude of the change in flux ratio is smaller
than the error bars on the STIS spectrum, justifying our decision not to vary K in
our main analysis.

\begin{figure} \includegraphics[width=0.5\textwidth]{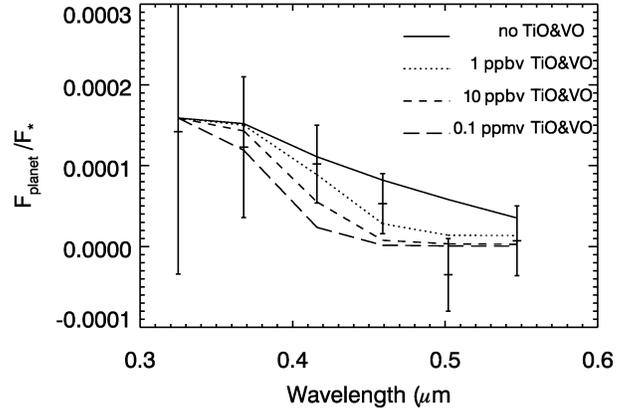} \caption{Synthetic, cloud-free
STIS spectra including TiO and VO. The presence of TiO and VO is unlikely in HD
189733b as it is expected to be too cold for TiO and VO condensates to evaporate, but
including these species in the model produces a good fit to the spectrum.
\label{tiovo}} \end{figure}

HD 189733b is not expected to have a hot enough photosphere for absorption due to
gaseous TiO and VO to be present in spectra \citep{fortney10}, and no
features due to these gases have been observed in the transmission spectrum; however, it is still
worthwhile testing the effect of these gases on the spectrum should they unexpectedly
be present. We show a series of cloud-free synthetic spectra including absorption due
to TiO and VO (using absorption data from \citealt{freedman11}) in Figure~\ref{tiovo}. The effect is much more significant than the
effect of varying the K abundance, and including some TiO and VO as well as Na (10
ppb) improves the spectral fit from the case without TiO and VO. This should not be
taken as evidence that TiO and VO are present on HD 189733b, as the problem is
degenerate and we still expect this planet to be too cold for these species to be
present in the gas phase; in addition, TiO and VO have not yet been detected in the
atmospheres of hotter planets (e.g. \citealt{sing13}), so the presence of these
molecules in planetary atmospheres is questionable.

\subsection{Cloud composition} \label{composition} 
\begin{figure*} \centering
\includegraphics[width=0.75\textwidth]{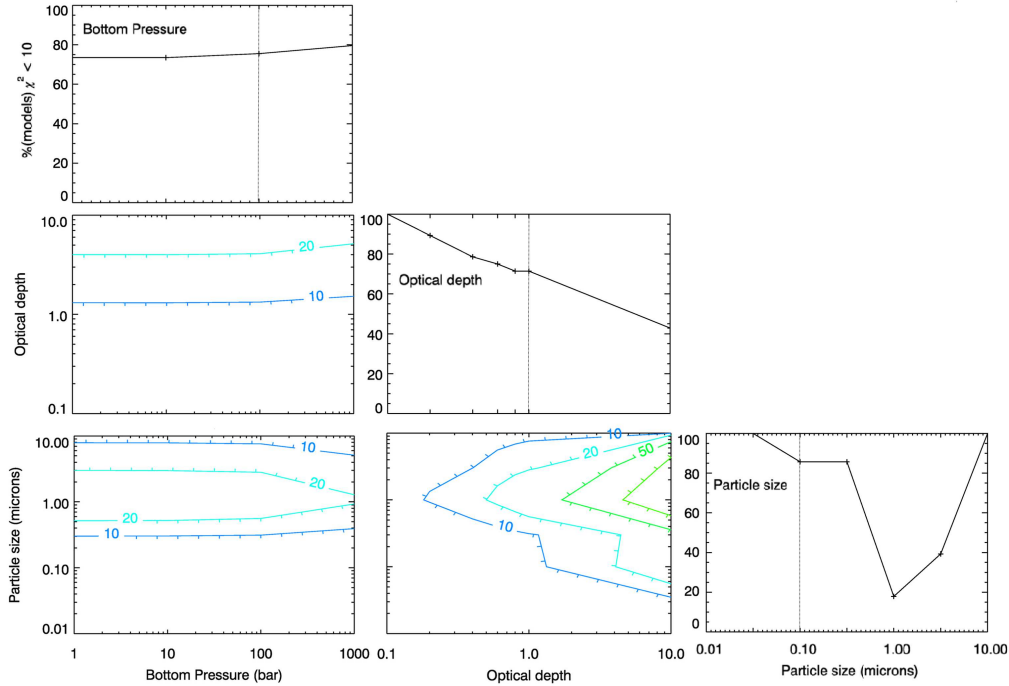} \caption{Contour
plots of $\chi^2$ goodness-of-fit parameter for cloud bottom pressure, optical depth
and particle size for MnS clouds. Also plotted are the percentages of models with
$\chi^2$ of $<$10 for each parameter. The dashed lines indicate the values used for
each parameter when held fixed. \label{mnscloud}} \end{figure*}

We have so far assumed that any
cloud on HD 189733b is made of enstatite. \citet{morley12} propose a range of other
minerals that are likely to form clouds on brown dwarfs, and MnS is a possible
candidate for brown dwarfs with similar temperatures to HD 189733b. We repeat the
analysis for a sodium VMR of 5 ppmv, with clouds composed of MnS. The results are
presented in Figure~\ref{mnscloud}.

It is clear that the particle size/optical depth parameter space still encompasses
the most important spectral variability for the cloud, as discussed in the Appendix of \citet{lee13a}. However, choosing MnS instead of
enstatite results in a greater range of cloud models that produce a fit
to the data with $\chi^2<$10, but no models that provide a very good fit to the data
($\chi^2<$5). This is due to the fact that MnS has a different refractive index to
enstatite, which affects the scattering efficiency and phase function of the
particles, and serves to illustrate the complexity of the cloud
parameter space. 

\subsection{Comparison with terminator cloud models} The results presented here
illustrate the highly degenerate nature of any cloudy atmosphere solutions for HD
189733b. If we are to assume that the dayside atmosphere is similar to the terminator
atmosphere, then we can place some further constraint on the cloud by comparing with
results from the transmission spectrum. The analysis of \citet{lee13} shows that the
best-fit models for the terminator atmosphere, for a uniformly-distributed cloud,
have particles $\sim$0.1 $\upmu$m in size, with an optical depth of 0.01 at 1.0
$\upmu$m. This corresponds to our set of solutions for uniformly-distributed 0.1
$\upmu$m particles with an optical depth of 0.5 at 0.25 $\upmu$m, as the extinction
cross-section of 0.1 $\upmu$m enstatite particles is a factor $\sim$50 smaller at 1.0
$\upmu$m than at 0.25 $\upmu$m. A $\chi^2$ of $<$6 is achieved for these particles if
the Na VMR is 5 ppmv, and $\chi^2$ $<$3 for a VMR of 50 ppmv. If we can expect the
cloud structure to be the same at the terminator and on the dayside, we could
therefore place some limits on the Na VMR.

Whether or not we expect the same cloud to form is a question for general circulation
models of HD 189733b to solve. So far, few have dealt with cloudy atmospheres, with
some exceptions. \citet{parmentier13} present a 3D circulation model for the somewhat
hotter planet HD 209458b, including passive tracer particles of various sizes to
examine the relative effects of circulation and sedimentation on aerosols and
condensates; they find that, whereas larger particles sediment out and show
significant longitudinal variability in abundance, the number of 0.1 $\upmu$m-sized
tracer particles remains relatively uniform with longitude. This indicates that a
cloud composed of 0.1 $\upmu$m particles could easily be approximately uniform
throughout the atmosphere of a hot Jupiter, and therefore we may be observing the
same cloud deck on the dayside as \citet{lee13} observe at the
terminator, although it is likely that different altitudes are
probed.

\begin{figure} \centering \includegraphics[width=0.5\textwidth]{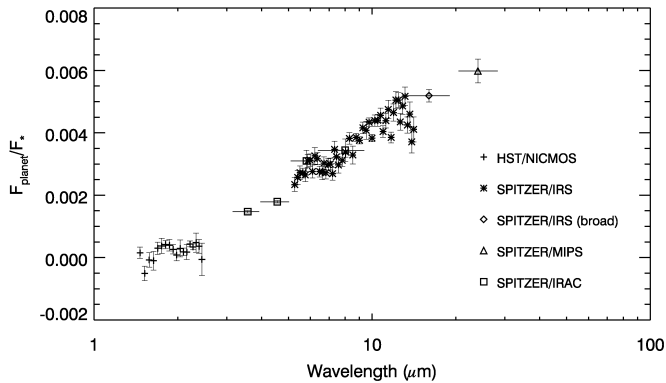}
\caption{The data for the thermal emission retrieval. NICMOS points are as reanalysed
in this work; 3.6 $\upmu$m and 4.5 $\upmu$m IRAC points are from \citet{knutson12};
the IRS spectrum is from \citet{grillmair08}; the IRS broadband point was originally
measured by \citet{deming06}; 5.8 and 8.0 $\upmu$m\textit{Spitzer}/IRAC and MIPS
points are as measured by \citet{charb08}, along with a reanalysis of the broadband
IRS point.\label{spectrum}} \end{figure}

\section{Thermal emission retrieval} We now take a subset of the cloud models used in
Section~\ref{cloudmodels} and examine their effect on retrievals of temperature and
molecular abundances from the infrared secondary eclipse spectrum
(Figure~\ref{spectrum}). The retrieval is performed as in
\citet{lee12} using hemispheric integration for non-scattering runs and a
45$^{\circ}$ emission angle approximation for scattering runs. We also
use our reanalysed NICMOS spectrum and the warm \textit{Spitzer} 3.6,
4.5 and 8 $\upmu$m
points presented in \citet{knutson12}. We include the cloud model in the model
atmosphere, but do not make any attempt to retrieve it; instead, we examine its
effect on the other parameters in the retrieval.

\begin{figure*} \centering \includegraphics[width=0.85\textwidth]{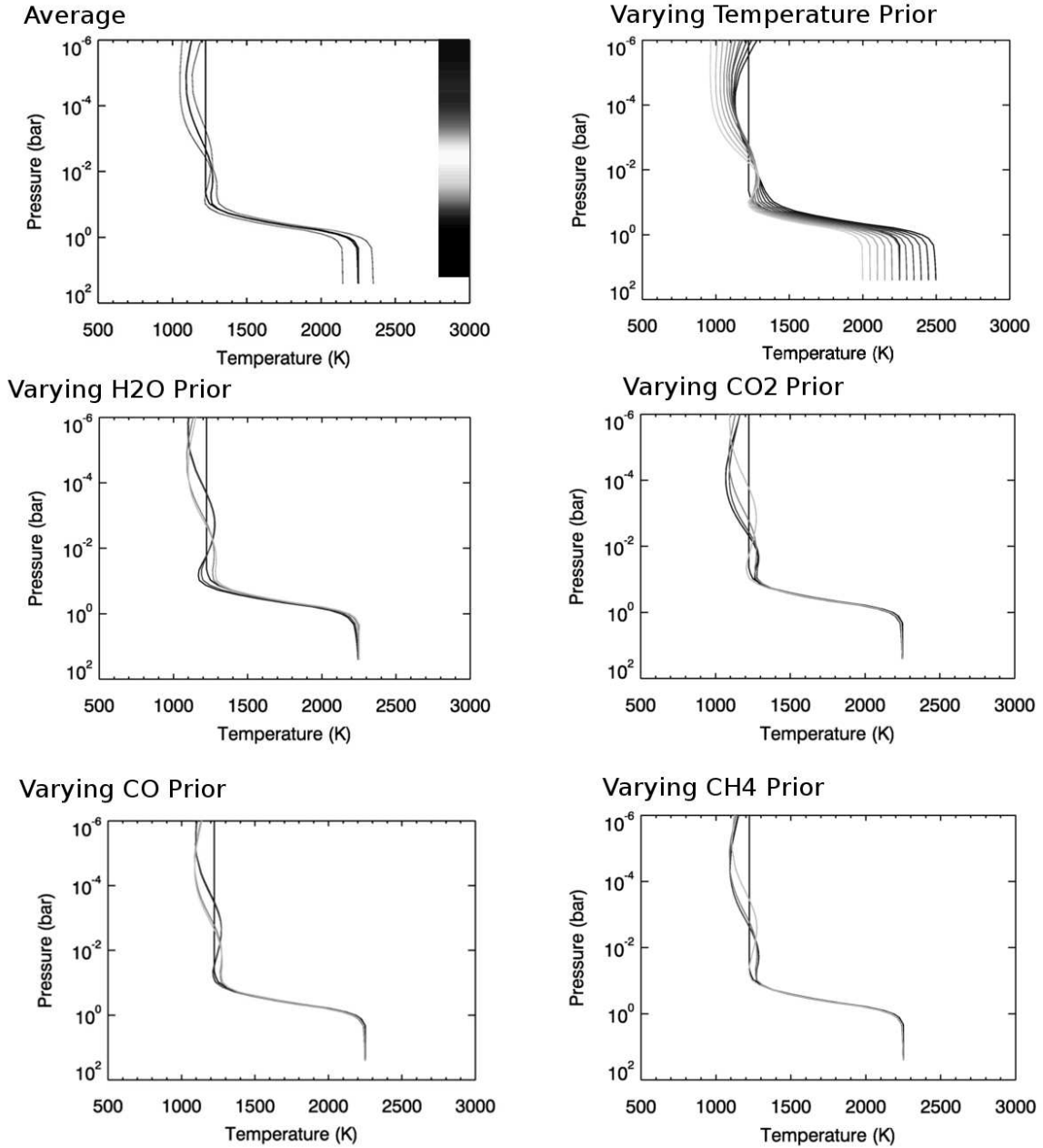}
\caption{Retrieved temperature profiles for a range of different
  priors. The mean and standard deviation are shown in the top left; the shaded bar
shows the average sensitivity to temperature at each altitude, across all
wavelengths, with black being low sensitivity and white high sensitivity. The maximum
sensitivity occurs between 1 and 100 mbar. The varying temperature
prior is
the best-fit case from \citet{lee12b}, shifted at 50 K intervals from -250K
(lightest grey line) to +250K (darkest grey line), with a prior error
of $\pm$50 K. We also vary the prior altitude-independent gas VMR for H$_2$O, CO$_2$, CO and CH$_4$.
The darkest---lightest grey colours correspond to priors for each gas of 1 ppmv, 10
ppmv, 100 ppmv and 1000 ppmv. It can be seen that the greatest
degeneracy occurs for  CO$_2$. The temperature prior used where the
gas priors are varying is the thick black line, with a prior error
of $\pm$50 K.\label{cloudfreetemp}} \end{figure*}

Firstly, we repeat the cloud-free analysis of \citet{lee12} with the
addition of the reanalysed NICMOS data
points; \citet{lee12} state that the NICMOS spectrum provides information about the
temperature profile in the deep atmosphere, and also the abundances of trace gases,
and as these data have changed we expect to retrieve a different temperature profile
and abundances. We perform the retrieval of temperature as a function
of pressure, over 50 atmospheric levels with a correlation length of
1.5 in ln(pressure), and
altitude-independent abundances of H$_2$O, CO$_2$, CO and CH$_4$ using a range of different temperature
and gas abundance priors; since the problem is underconstrained, it is necessary to
test the influence of the prior on the retrieved property. We present the results of
this analysis for temperature in Figure ~\ref{cloudfreetemp} and
Table~\ref{cloudfreegases}. The best-fit retrieved temperature profile
(Figure~\ref{cloudfreetemp} top left) and gas abundances are averaged over all runs. The
temperature sensitivity weighting function averaged over all wavelengths is shown
qualitatively in Figure~\ref{cloudfreetemp} by a shaded bar, with the
peak (lightest colour)
occuring between 1 and 100 mbar; this corresponds to the region of the atmosphere
probed by the majority of the data, and therefore the temperature is least dependent
on the temperature prior in this region; this is shown in
Figure~\ref{cloudfreetemp} top right, in which the retrieved profiles for different priors
converge between 1 and 100 mbar. 

The temperature structure is most
sensitive to the \citet{knutson12} \textit{Spitzer}/IRAC 3.6 and 4.5 $\upmu$m points as
these data have by far the smallest fractional error, meaning
that they drive the temperature retrieval. This accounts for the
differences in the shape of the retrieved T-p profile compared with that of
\citet{lee12}. The 3.6 $\upmu$m weighting function peaks at around 1
mbar and constrains the stratospheric temperature, whereas the
4.5 $\upmu$m weighting function peaks at the tropopause, at around 100
mbar. 

\begin{table} \centering \begin{tabular}[c]{|c|c|c|c|c|} \hline & Mean$\pm\sigma\times10^{-4}$
& Min& Max & Lee et al.\\ \hline H$_2$O&
11$\pm$11 & 1.9 &46 & 6.7\\ CO$_2$& 22$\pm$28 &1.3& 96& 20\\
CO&2.1$\pm$4.1  & 0.010 & 22 & 29\\ CH$_4$&0.26$\pm$0.09 & 0.01 & 0.42
& 0.0019\\
\hline \end{tabular} \caption{The mean retrieved values for each gas VMR, and the
minimum and maximum retrieved values over all priors. H$_2$O and CH$_4$ can be
reasonably well constrained, but CO$_2$ and CO have $\sigma$ larger than 100\% of the
mean value. The best-fit values for \citet{lee12} are given for
comparison.\label{cloudfreegases}} \end{table}

In Figure~\ref{cloudfreetemp} it can be seen that the greatest degeneracy
between temperature and gas abundance occurs for H$_2$O and CO$_2$,
with the precise shape of the retrieved T-p profile between 1 and 100
mbar varying as a function of the H$_2$O and CO$_2$ abundance
priors. This is because the peak of the temperature weighting function and therefore
the pressure probed varies with atmospheric opacity, and these two gases are the most
spectrally active. The increase in the error bars for the NICMOS spectrum means that
the gas abundance retrievals are more sensitive to the prior than found by
\citet{lee12}, so for a small H$_2$O or CO$_2$ prior abundance the retrieved
abundance is smaller, the atmospheric opacity is less and the pressure probed is
higher. The
reverse is true for high prior abundances of H$_2$O or CO$_2$. The effects of
changing the CO and CH$_4$ abundances are smaller, but similar.  

The average temperature profile retrieved is largely similar to that
of \citet{line13}, except that we see a slight decrease in temperature
above the 1 mbar level compared with \citet{line13}. This is probably
due to the fact that we retrieve temperature for each atmospheric
level individually, with an assumed correlation length (1.5 in
log-pressure coordinates), whereas \citet{line13} parameterise the
temperature profile. Therefore, the upper atmosphere temperature
retrieved by \citet{line13}
cannot vary freely with respect to the temperature between 1 and 100
mbar, where the sensitivity is greatest, whilst we retrieve a
continuous profile. 

The mean, standard deviation and range of retrieved gas VMRs are given in
Table~\ref{cloudfreegases}. The standard deviations are highest for CO$_2$ and CO,
implying a high dependence on the prior. The retrieved values for
H$_2$O and CO$_2$ are similar to
that of \citet{lee12}, but the value for CO is smaller, although due to
the dependence on the prior this result must be viewed with caution. However, we
retrieve a much higher CH$_4$ abundance than \citet{lee12}, roughly a factor of 10
higher than the \textit{Spitzer}-derived upper limit of
\citet{madhu09}. Our higher value for CH$_4$ is similar to the result of
\citet{line13},  and so it is likely to be a result
of the revised 3.6 and 4.5 $\upmu$m points which are also used by
these authors.

We now seek to investigate the effect of cloud on the retrieved temperature profile
and gas abundances. We use the average temperature prior and \textit{a priori} gas
VMRs of 10$^{-4}$ for all retrieved species. The parameters for the set of models we
use are shown in Table~\ref{cloud_mods_emission}. These correspond to some of the
best-fitting solutions from the previous analysis that have very different infrared
optical depths, so are likely to affect the thermal emission retrieval differently.

\begin{table} \centering \begin{tabular}[c]{|c|c|c|c|} \hline Model & Base p &
0.25 $\upmu$m O. D. & Particle radius \\ \hline 1 & Uniform & 10 & 0.1
$\upmu$m\\
2& 1000 mbar& 1 & 0.01 $\upmu$m\\ 3& 1000 mbar& 1 & 3 $\upmu$m\\ 4& 10 mbar& 10
& 0.03 $\upmu$m\\ \hline \end{tabular}
\caption{The parameters for the cloud models included in thermal emission retrievals,
to test the dependence on cloud properties.\label{cloud_mods_emission}} \end{table}

\begin{figure*} \centering \includegraphics[width=1.0\textwidth]{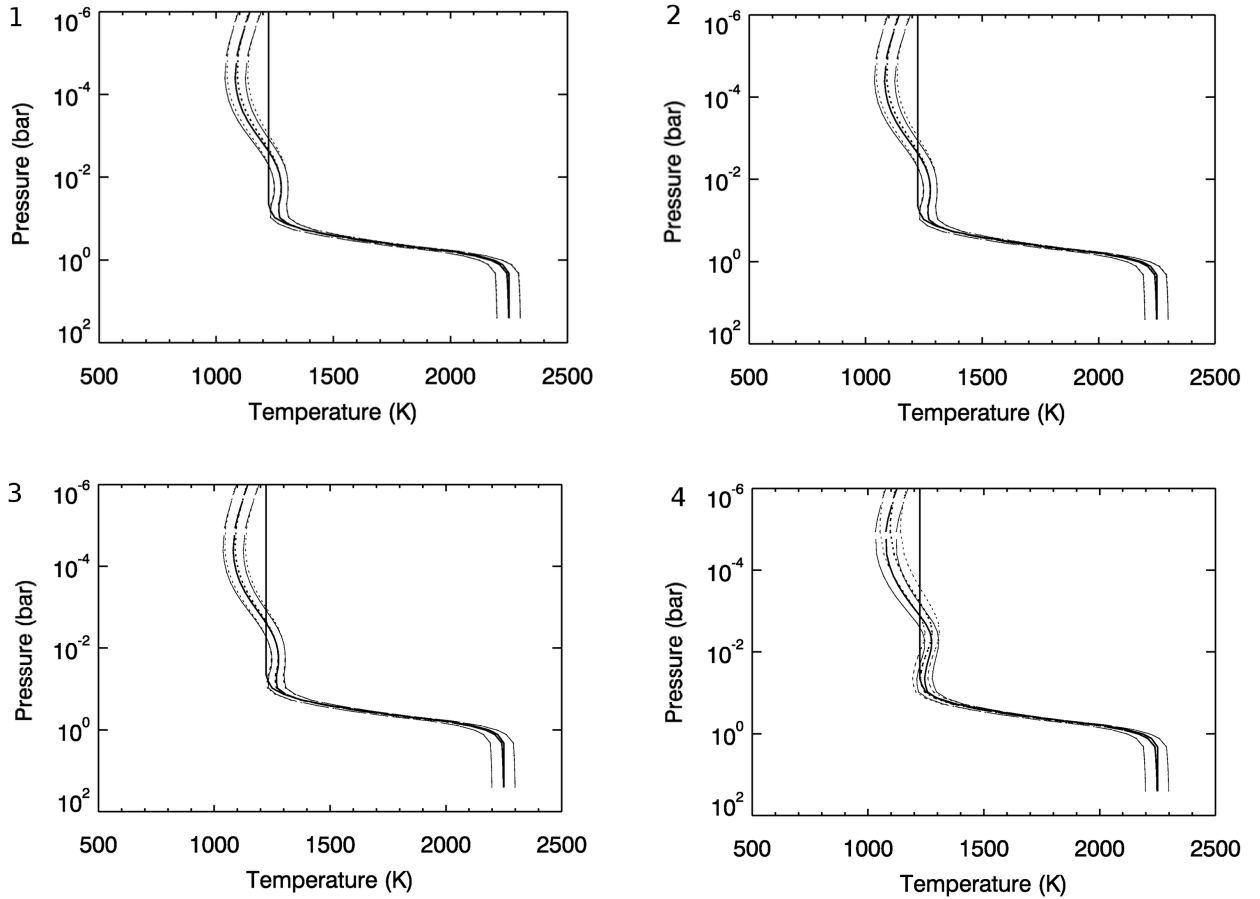}
\caption{Retrieved temperature profiles for each cloud model. Thick solid lines show
the \textit{a priori} temperature, dotted lines the extinction-only retrieval and
dashed lines the multiple scattering retrieval. Thick lines show the retrieved value
and thin lines show the error envelope.\label{retrievedtemp}} \end{figure*}

We show the retrieved temperature profiles and gas VMRs for each of the cloud models,
as well as the gas abundances for the cloud-free case (Figure~\ref{retrievedtemp},
Table~\ref{retrieved_gas}). We perform each of the retrievals twice, first using the
fast, extinction-only disc integration technique as in \citet{lee12}, and secondly
using multiple scattering as in Section~\ref{cloudmodels}, except we approximate the
disc with a single 45$^{\circ}$ emission/incidence angle as the 5-angle calculation
is extremely time-consuming for a full retrieval.

\begin{table*} \begin{minipage}{126mm} \centering \begin{tabular}[c]{|c|c|c|c|c|}
\hline Model & H$_2$O VMR (10$^{-4}$)& CO$_2$ VMR (10$^{-4}$)& CO VMR (10$^{-4}$) &
CH$_4$ VMR (10$^{-4}$)\\ \hline 1 & 9.7$\pm$4.9/11$\pm$6
&8.9$\pm$4.6/8.1$\pm$4.2&1.4$\pm$1.4/1.4$\pm$1.4 &0.25$\pm$0.11/0.26$\pm$0.11\\ 2&
9.6$\pm$4.9/11$\pm$6&9.0$\pm$4.7/8.2$\pm$4.3&1.4$\pm$1.4/1.4$\pm$1.4&0.25$\pm$
0.11/0.26$\pm$0.11\\ 3& 8.4$\pm$4.4/11$\pm$5 &8.1$\pm$4.2/7.8$\pm$4.1
&1.4$\pm$1.4/1.4$\pm$1.4& 0.24$\pm$0.10/0.25$\pm$0.11\\ 4&
3.8$\pm$2.0/6.8$\pm$3.7&31$\pm$17/13$\pm$7 &1.3$\pm$1.3/1.3$\pm$1.3&0.31$\pm$
0.09/0.31$\pm$ 0.11\\ 5& 9.7$\pm$4.9&8.9$\pm$4.6&1.4$\pm$1.4&0.25$\pm$0.11\\ \hline
\end{tabular} \caption{The retrieved gas VMRs for each of the four cloud models,
including the cloud-free case (model 5), for extinction-only (left) and multiple
scattering (right). The cloud-free case was modelled using an extinction-only
approximation. \label{retrieved_gas}} \end{minipage} \end{table*}

It is clear from Figure \ref{retrievedtemp} and
Table~\ref{retrieved_gas} that the thermal emission retrieval is largely insensitive
to the inclusion of clouds corresponding to our best-fitting models. This is due to
the fact that the majority of best-fitting models with optical depths greater than 1
at 0.25 $\upmu$m have small particle sizes, and therefore the extinction drops
rapidly as a function of wavelength. These particles have little
effect at wavelengths longer than 1 $\upmu$m because the particle size is small with respect to the wavelength of the
light, and so the optical depth in the infrared is negligible even though it is
significant at shorter wavelengths. This means that even higher optical depths of 0.1
$\upmu$m-sized particles would be unlikely to change the temperature retrieval.

\begin{figure} \includegraphics[width=0.5\textwidth]{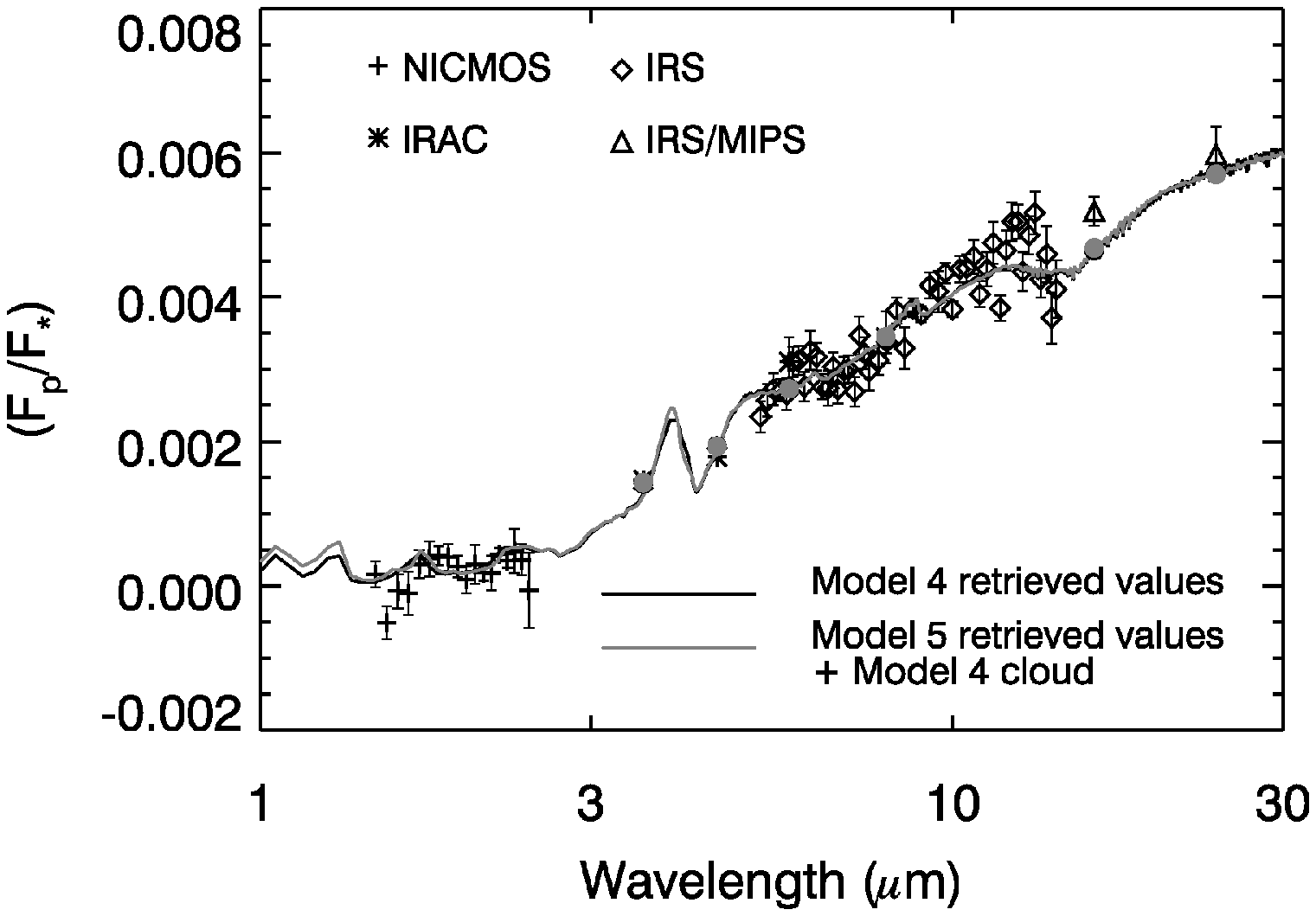} \caption{We
show synthetic spectral fits for a forward model containing the cloud from model 4
but the retrieved parameters from model 5 (grey) and the retrieved parameters and
cloud from model 4 (black). The difference between the two spectra is
small, with $\chi^2$ of 142 and 138 respectively. The numbers of
spectral points (71) and degrees of freedom (54) are the same for both cases. The circles show the high-resolution synthetic spectra
convolved at the \textit{Spitzer} IRAC, broadband IRS and MIPS bandpasses.\label{0_03_spectra}} \end{figure}

The exception is the retrieved H$_2$O and CO$_2$ abundances and temperature profile for cloud
model 4, using an extinction-only assumption; including a high (10---1 mbar) cloud
of 0.03 $\upmu$m-sized particles results in a larger retrieved CO$_2$ abundance,
which in turn affects the shape of the temperature profile in this pressure range (as
in Figure~\ref{cloudfreetemp}). Where solutions are degenerate, as in this case, it
is possible for small changes in one model variable (here, the cloud optical depth)
to result in different best-fit solutions being obtained. In
Figure~\ref{0_03_spectra}, we show the synthetic spectrum for the model 4 retrieved
case, and the spectrum for a forward model with retrieved temperature profile and gas
abundances as for a cloud-free model but including a model 4 cloud. The spectra are very
similar, with $\chi^2$ of 138 and 142 respectively, demonstrating that the difference
in retrieved CO$_2$ abundance for this case should not be treated as significant.
Additionally, the multiple scattering retrieval for model 4 yields a lower abundance
of CO$_2$ and a temperature profile more similar to those retrieved using other cloud
models. For the case of HD 189733b, therefore, it seems that an accurate retrieval of
temperature and atmospheric composition from the thermal emission spectrum may not be
critically dependent on an understanding of the cloud properties.

\section{Discussion} We have used a very simple set of cloud models to
investigate the acceptable parameter space for the HD 189733b STIS spectrum. The
quality of the data at this stage do not warrant more detailed modelling, but it is
instructive to consider the most important steps that could be taken to improve this
model in anticipation of future results.

\subsection{Monodisperse particles} We have assumed that all the particles in the
cloud are of a single size. This is of course not likely to be the case in practice;
cloud models based on microphysical processes indicate that particle sizes will
change as a function of altitude, and at a given altitude particle size distributions
may be relatively broad, for example as discussed by \citet{helling08}. Coalescence
processes, condensational growth and sedimentation will result in the presence
of larger grains towards the bottom of the cloud deck.

Broadening the size distribution will make the cloud optical depth more uniform as a
function of wavelength; extinction due to cloud particles is most efficient when the
particle size is comparable to the wavelength of light. This may result in the cloud
having a larger effect on the thermal emission spectrum than we find in this work;
although in most cases we find that particle sizes of larger than 0.1 $\upmu$m do not
provide a good fit to the data at short wavelengths, this does not preclude larger
particles being present as they would scatter the short wavelength light less
strongly than smaller particles.

When the data are available that would allow us to test a more complex model, we
consider that investigating different particle size distributions would be the most
important step to take, as particle size is the property that has the largest effect
on the spectrum.

\subsection{Particle composition} The composition of clouds on cool brown dwarfs and
hot exoplanets is treated somewhat differently in different models. \citet{morley12}
consider a range of possible cloud compositions for specific temperatures, under the
assumption that if clouds of more than one species are present they will form at
different temperatures, and therefore altitudes. \citet{helling08} have a very
different model, in which multiple species condense onto the same TiO$_2$ `seed
particle' \citep{helling06}, thus forming grains of mixed composition. This
significantly complicates the calculation of spectral properties for these clouds;
however, \citet{helling08} find that, for a low-gravity atmosphere with
T$_{eff}\sim$1300 K, the enstatite is the constituent with the largest effect on the
spectrum, meaning that enstatite probably represents the best approximation to the
multi-component grains.

We tested the effect of including MnS, as proposed by \citet{morley12} for cool brown
dwarfs, and find that the different spectral properties of the material do affect the
spectrum, but to a lesser extent than the size of the particles. Therefore, particle
composition is an interesting route for further investigation, but is of secondary
importance to developing more detailed representations of particle size.

\subsection{Effect of temperature profile} 
\begin{figure} \includegraphics[width=0.5\textwidth]{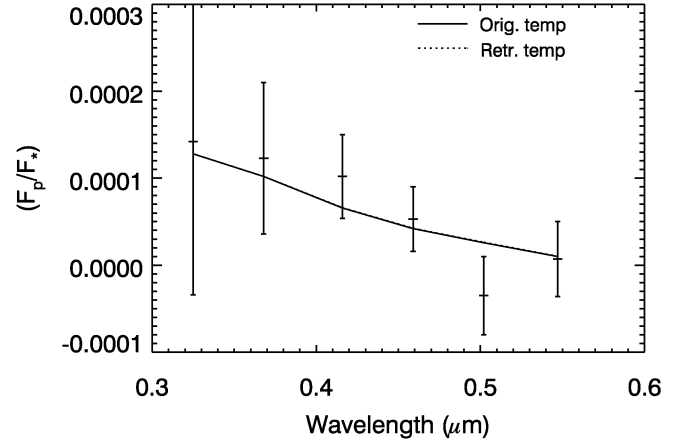} \caption{We
show two STIS spectrum models, both containing the model 1 cloud
properties. One is the model generated using the original temperature
profile, the other uses the retrieved temperature profile for model
1. The synthetic spectra are indistinguishable from each other.\label{spectra_temptest}} \end{figure}

We have modelled the STIS reflection
spectrum using the best-fit temperature profile from \citet{lee12b}, but have then
retrieved a different profile using the thermal emission spectrum. We tested the
effect on the best-fit cloud properties of including the altered temperature profile
in our analysis of the cloud. We found that the changes to the spectral shape were
negligible compared with the error bars on the STIS spectrum (Figure~\ref{spectra_temptest}), and the $\chi^2$ values for
different cloud models do not change greatly. Therefore small inaccuracies in the
temperature profile we use would not significantly affect our conclusions; the same
cloud models provide good/poor fits to the measured spectrum regardless of the
temperature profile used.

The reason for the limited effect of temperature is likely to be that the absorption
due to Na at the altitudes to which we are sensitive is dominated by
pressure-broadened line wings. Temperature variation is more likely to affect the line
core absorption at higher altitudes, but at the resolution of the STIS spectrum such
variation would not be seen. The STIS measurements also lie in a wavelength region
where the signal is dominated by reflected starlight, rather than light emitted from
the planet.

\section{Conclusion} We find that a large range of enstatite cloud models can fit the
measured \textit{HST}/STIS spectrum. Cloud-free model atmospheres are also acceptable
solutions, although we consider this to be less plausible due to the clear evidence
from the transmission spectrum that HD 189733b is cloudy \citep{pont13,lee13} and the
likelihood that small particles are evenly distributed in hot Jupiter atmospheres
\citep{parmentier13}. Small enstatite particles ($<$0.1 $\upmu$m) and 50 ppmv of Na
provide the best fit to the STIS data of the examples
  we test; we find an overlap with the models of
\citet{lee13} for a uniformly-distributed 0.1 $\upmu$m cloud with an optical depth of
0.5 at 0.25 $\upmu$m. However, the problem is extremely degenerate and we cannot
exclude solutions with larger cloud particles.

The retrieval of temperature and atmospheric composition from the thermal emission
spectrum is relatively insensitive to the inclusion of cloud in the model atmosphere,
for our best-fitting models. This suggests that for the case of HD 189733b accurate
retrieval of temperature and gaseous abundances from the thermal emission spectrum is
possible, even without detailed knowledge of the cloud properties. Solution
degeneracy prevents firm conclusions from being drawn about the nature of the cloud
on HD 189733b; the current quality and coverage of spectroscopic data for most
exoplanets is therefore insufficient to simultaneously constrain temperature
structure, gaseous abundances and multiple cloud properties. Given the additional
complexity (and therefore number of degenerate solutions) introduced to the retrieval
problem when clouds are included, this implies that the best approach with the
currently available secondary eclipse data is to use cloud-free model atmospheres for
temperature retrieval. As data quality improves, alternative, more detailed cloud
models must be explored and their effect on the emission spectrum reassessed.

\section*{Acknowledgements} JKB acknowledges the support of the John Fell Oxford
University Press (OUP) Research Fund for this research and LNF is supported by a
Royal Society Research Fellowship. We thank the anonymous reviewer for
his/her report, and Tom Evans, Fr\'ed\'eric Pont, David Sing and Kevin
Heng for helpful discussions about the visible albedo observations. 

\bibliography{bibliography}

\label{lastpage} \end{document}